# Fabrication of (In,Ga)N pseudo-substrates by a three-step growth protocol without ex-situ processing

Huaide Zhang, Aidan F. Campbell, Jingxuan Kang, Jonas Lähnemann, Oliver Brandt, Lutz Geelhaar [a]

*Paul-Drude-Institut für Festkörperelektronik, Leibniz-Institut im Forschungsverbund Berlin e.V., Hausvogteiplatz 5-7, 10117 Berlin, Germany*

[a] Corresponding author: geelhaar@pdi-berlin.de

We fabricate (In,Ga)N pseudo-substrates with a total thickness of ≈1 μm grown on GaN templates using plasma-assisted molecular beam epitaxy. In a three-step process, we change growth conditions from N-rich to metal-rich in order to sequentially form a roughened GaN layer, relaxed (In,Ga)N nanostructures, and a coalesced, smooth (In,Ga)N layer. Samples are analyzed by scanning electron and atomic force microscopy, X-ray diffraction, as well as photo- and cathodoluminescence spectroscopy. Compared to a reference layer grown directly on GaN, the pseudo-substrate exhibits a higher In content (≈0.3), strain relaxation degree (≈80%), narrower photoluminescence linewidth, and larger area fraction of bright regions in cathodoluminescence maps, showing the benefits of the three-step growth protocol. This straightforward approach does not necessitate any ex-situ processing and could enable the scalable fabrication of (In,Ga)N pseudo-substrates for high-efficiency red-emitting (In,Ga)N devices.

The development of micro light-emitting diode (μLED) displays requires integrated red, green, and blue (RGB) emitters. Conventionally sized blue and green LEDs are based on (In,Ga)N and their red counterparts on (Al,Ga,In)P. All of them are very efficient, but device miniaturization typically degrades performance. Blue and green (In,Ga)N μLEDs retain an external quantum efficiency of 20-35%,[1,2] whereas the one of red (Al,Ga,In)P μLEDs drops to ≈10%.[3,4] The latter value is comparable to what can be achieved for red *(In,Ga)N* μLEDs.[5] Since the efficiency of LEDs based on this material is less sensitive to size reduction, research aimed at improving red (In,Ga)N LEDs recently intensified.[6]

Efficient red (In,Ga)N emission is challenging due to the large lattice mismatch between the quantum wells (QWs) and the GaN typically used as substrate. On the one hand, the high compressive strain hinders In incorporation, thus, achieving 30-40% In content necessary for red emission is difficult.[7] On the other hand, the strain gives rise to strong polarization fields. The resulting quantum-confined Stark effect spatially separates carriers, thus reducing radiative recombination and internal quantum efficiency.[8,9] Additionally, strain induces initially point and eventually structural defects,[10–12] both of which may act as non-radiative recombination centers.

The root cause for these challenges is the large lattice mismatch, which would be reduced if the LED structure were grown on a relaxed (In,Ga)N pseudo-substrate instead of GaN. In analogy to blue (In,Ga)N LEDs with about 15% In[13] in QWs grown on GaN templates, for red LEDs a fully relaxed (In,Ga)N layer with 15-25% In would be desirable as pseudo-substrate, corresponding to an in-plane lattice constant of 3.24-3.27 Å. Various strategies have been pursued to fabricate such pseudo-substrates for the growth of red LEDs. Partial strain release has been achieved by porosification via (electro)chemical etching but this approach suffers from rough surface with a high density of pits.[14–16] As another strategy, (In,Ga)N layers are transferred to a different carrier substrate, a process coined InGaNOS, but the level of strain relaxation is limited.[17] All of these efforts involve intricate processing steps that increase fabrication complexity and cost.

In this work, we present a straightforward three-step growth protocol to fabricate relaxed (In,Ga)N pseudo-substrates without the need of ex-situ processing. The underlying idea is to adjust the growth conditions in plasma-assisted molecular beam epitaxy (PAMBE) in such a way that initially a nanostructured surface forms that is subsequently smoothened. The growth sequence is illustrated by the schematics in Fig. 1(a) and 1(a'). As step 1, GaN is grown under N-rich conditions known to produce a rough surface with pits.[18–20] In step 2, In is added and the In content is increased. The continued N-rich environment is expected to lead to columnar growth.[21,22] Our key achievement is the (In,Ga)N coalescence and surface smoothening under metal-rich conditions in step 3.

Samples were grown by PAMBE on commercial GaN templates. For step 1, the substrate temperature was 550 °C, the growth duration 30 min, and the N and Ga flux were $6.8 \times 10^{14}$ and $3.1 \times 10^{14}$ atom/(s·cm$^2$), respectively. For step 2, the In shutter was opened and its effusion cell temperature was intentionally ramped up gradually over 15 min from 750 °C to about 850 °C until an expected In flux of $\approx 1.7 \times 10^{14}$ atom/(s·cm$^2$) was reached, while the N and Ga flux were maintained constant. This resulted in an In/(In+Ga) flux ratio of $\approx 0.35$, and (In+Ga)/N flux ratio of $\approx 0.7$. These conditions were then kept for a further 15 min to stabilize the In content in the alloy. For step 3, the Ga and In flux were adjusted to $4.7 \times 10^{14}$ and $3.4 \times 10^{14}$ atom/(s·cm$^2$), respectively [Ga/N flux ratio $\approx 0.7$ and (Ga+In)/N flux ratio $\approx 1.2$], and the growth time was 60 min.

Four types of samples were fabricated. Sample #A was stopped after step 2. In a separate growth experiment, sample #B and a reference sample #R were grown side by side by starting directly with step 3 on a piece of sample #A and a fresh GaN template, respectively. For sample #C, all three steps were executed subsequently without any interruption.

The samples were characterized using a suite of techniques to assess their structural and optical properties. Surface morphology was monitored in-situ by reflection high-energy electron diffraction (RHEED) and examined ex-situ using scanning electron microscopy (SEM). The root mean square (RMS) surface roughness was quantified by atomic force microscopy (AFM). A triple-axis high resolution x-ray diffractometer (Philips PANalytical X'Pert PRO MRD) equipped with a two-bounce hybrid Ge(220) monochromator and a CuKα1 source (λ = 1.540 598 Å) was employed to determine the lattice constant, In content, and strain relaxation degree based on $2\theta$-$\omega$ scans of the 0002 and $10\bar{1}5$ reflections as well as reciprocal space maps (RSMs) around the $10\bar{1}5$ reflection.[23,24] Photoluminescence (PL) spectra were acquired at room temperature (RT) using a Renishaw InVia setup to investigate the emission properties. The samples were excited by a 325 nm He-Cd laser with a spot diameter of about 1 μm and an excitation density of

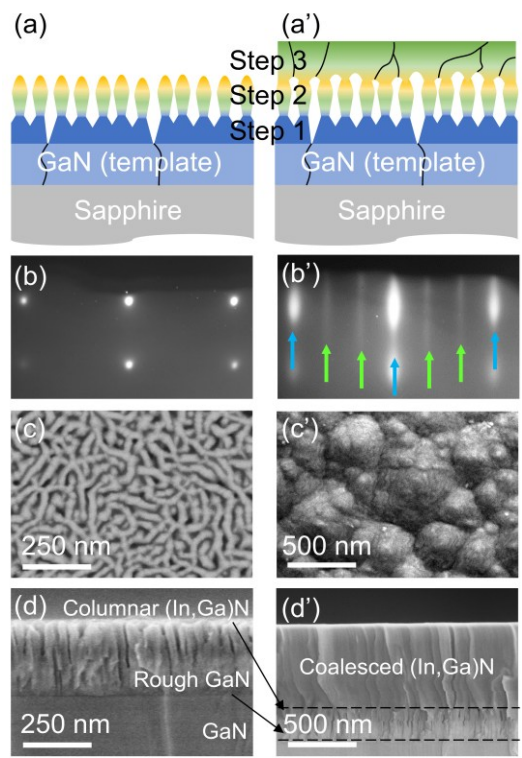

Fig. 1: Morphological and structural evolution of the pseudo-substrate during the three-step growth process. The left column (a, b, c, d) corresponds to sample #A terminated after step 2, while the right column (a', b', c', d') shows the final state after step 3 (sample #B). Schematic illustrations of (a) the columnar nanostructure before coalescence and (a') the final, smooth pseudo-substrate. Representative RHEED patterns observed along the [$10\bar{1}0$] azimuth, showing the transition from (b) a pure transmission pattern to (b') a predominantly reflection pattern. (c, c') Top-view and (d, d') cross-sectional SE micrographs.

≈1 kW cm$^{-2}$. The spatial distribution of light emission was analyzed by panchromatic cathodoluminescence (CL) maps recorded at RT.

The initial GaN layer grown under N-rich conditions (step 1) exhibits a transmission RHEED pattern, which is similar to the one in Fig. 1(b), indicative of a rough surface morphology. The presented RHEED pattern belongs to the (In,Ga)N layer grown as the second step (sample #A) and also reveals a rough morphology.[25] Indeed, the top-view and cross-sectional secondary-electron (SE) micrographs of this sample depicted in Fig. 1(c) and 1(d) exhibit a columnar, "brain-like" 3D surface structure of winding nanowalls, as expected from similar studies.[21,22] The formation of these nanostructures can be explained by two phenomena. First, it is well-known that N-rich conditions during GaN growth lead to insufficient adatom mobility and, hence, a morphology characterized by mounds, trenches and pits.[18,20] Second, the lattice mismatch between (In,Ga)N and GaN leads to preferential In incorporation at the top of the GaN mounds,[26–28] where strain can be elastically relaxed most efficiently. Thus, the columnar nanostructures are expected to be highly relaxed. In addition, threading dislocations likely bend to the free sidewalls and terminate there, analogous to the situation in bottom-up grown nanowires.[29]

A statistical analysis of the micrographs indicates for the nanowalls a mean width of approximately 30 nm and separating gaps of 20 nm. The thickness of the rough GaN layer and the columnar (In,Ga)N layer amount to approximately 100 nm and 150 nm, respectively.

The transition to metal-rich conditions in step 3 is designed to promote adatom mobility[30] and facilitate the coalescence of the underlying nanostructures into a smooth, continuous layer. Monitoring via RHEED shows the evolution from a transmission (spotty) [Fig. 1(b)] to a predominantly reflection (semi-streaky) pattern [Fig. 1(b')], indicating the desired transition from a rough 3D surface to a relatively smooth one. Furthermore, the appearance of a $(\sqrt{3} \times \sqrt{3})R30°$ surface reconstruction in the reflection pattern [green arrows in Fig. 1(b')] is characteristic for an In-terminated surface.[12,31,32] SEM analysis of sample #B reveals a comparatively smooth, fully closed surface [Fig. 1(c')], and a final layer thickness of approximately 700 nm [Fig. 1(d')]. These results demonstrate that it is possible to transition from the rough, columnar surface resulting from N-rich growth to a completely coalesced, smooth surface by metal-rich growth.

The structural properties of the pseudo-substrate (sample #B) and the reference (sample #R) are assessed on the basis of the RSMs around the asymmetric $10\bar{1}5$ reflection presented in Fig. 2(a) and 2(b), respectively. In both RSMs, the (In,Ga)N reflection is located between the vertical solid line indicating pseudomorphic growth on the GaN template and the dashed line corresponding to fully relaxed (In,Ga)N. However, the (In,Ga)N reflection of sample #B lies closer to the dashed line. Furthermore, the width of the reflection along the $Q_Z$-axis is smaller, implying an enhanced In uniformity. This sample exhibits a strain relaxation degree of approximately 80% and an In content of 0.3, corresponding to an $a$-lattice constant of 3.27 Å. This value meets the requirement for overgrowth by a red (In,Ga)N LED explained in the introduction. The results for sample #R are 50% and 0.28, respectively, with an $a$-lattice constant of about 3.24 Å. These values are consistent with those calculated from the approximate lattice constants derived from the $2\theta$-$\omega$ scans of the 0002 and 10$\bar{1}$5 reflections shown in Fig. 2(c). Sample #A displays a broader diffraction profile with lower intensity within the angular range characteristic of (In,Ga)N (approximately 33-34° for the 0002 reflection and 99-104° for the 10$\bar{1}$5 reflection). This broadening can be attributed to an intentionally introduced gradient in In content along the growth direction, suggesting a potential slight lattice misalignment between the (In,Ga)N nanowalls and the underlying GaN template. These observations are consistent with the diffusive features observed in the RSM for Sample #A (see Fig. S1 in the supporting information). Nevertheless, the nanowalls are still single crystalline, as verified by the transmission RHEED pattern [cf. Fig. 1(b)].

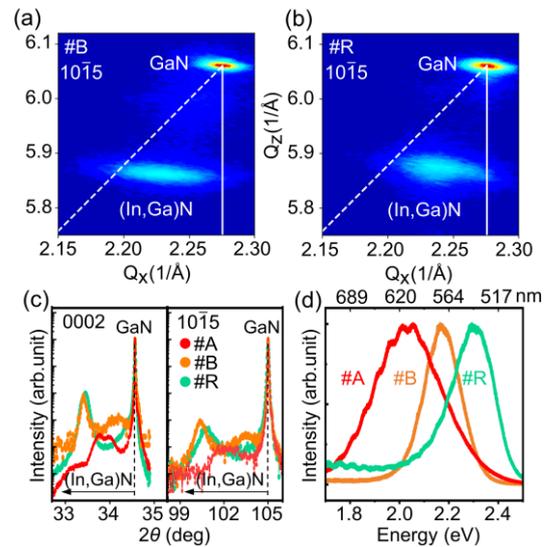

Fig. 2: Structural and optical characterization of different samples. RSMs around the asymmetric $10\bar{1}5$ reflection for (a) samples #B and (b) #R. The dashed white line corresponds to fully relaxed (In,Ga)N, while the solid white line indicates the fully strained condition coherent to the underlying GaN template. (c) Triple-axis with analyzer Ge (220) x-ray diffraction ($2\theta$-$\omega$) scans of the symmetric 0002 and asymmetric $10\bar{1}5$ reflections for the same samples. Peaks at 34.56° and 105.00° originate from the GaN template. The peaks at lower angles in both scans are from the (In,Ga)N layers. (d) RT-PL spectra of samples #A, #B, and #R.

The PL spectra of samples #A, #B, and #R are presented in Fig. 2(d). The peak wavelength for sample #A is about 610 nm, which is consistent with the In content of around 0.35 estimated from the In/(Ga+In) flux ratio in step 2. The PL band of sample #B peaks at 571 nm, representing a significant blue-shift compared to sample #A and a slightly smaller red-shift compared to the 541-nm peak of sample #R. These wavelengths are in line with the In content determined from the RSMs. Furthermore, the full width at half maximum (FWHM) of sample #B is smaller than that of both other samples (172 meV/45 nm vs. 195 meV/46 nm for sample #R and 311 meV/93 nm for sample #A). Moreover, sample #C grown under identical conditions as sample #B but without growth interruption shows a very similar PL peak wavelength of 576 nm with an even smaller FWHM of 140 meV/37 nm (see Fig. S2 in supporting information). We note that the In content and strain relaxation degree of sample #C are 0.32 and 75%, respectively (cf. RSM in Fig. S3 in supporting information), i.e. slightly different from sample #B. All details including XRD and PL measurements are listed in Table 1.

Table 1: Structural and optical properties of the different samples characterized by XRD and PL. $\lambda_{peak}$ is the peak wavelength of the PL band.

| Sample | | #A | #B | #C | #R |
|---|---|---|---|---|---|
| XRD | In | / | 0.30 | 0.32 | 0.28 |
| | R (%) | / | 80 | 75 | 50 |
| | $a$ (Å) | / | 3.27 | 3.27 | 3.24 |
| PL | $\lambda_{peak}$ (nm) | 610 | 571 | 576 | 541 |
| | FWHM (meV) | 311 | 172 | 140 | 195 |
| | FWHM (nm) | 93 | 45 | 37 | 46 |

The red-shift in the PL spectrum of sample #A compared to sample #B is consistent with the higher In content and full strain relaxation expected for the former sample. Likewise, the red-shift in the PL spectrum of sample #B compared to sample #R is a consequence of the higher In content and strain relaxation degree of the former sample. This difference between the two samples is very interesting because it demonstrates that the rough, nanostructured morphology and increased lattice constant after step 2 leads to enhanced In incorporation and strain relaxation compared to direct growth on a GaN template. Moreover, the narrower PL linewidth of the pseudo-substrates, samples #B and #C, indicates a substantial reduction in alloy disorder. This holds in particular for sample #C grown without interruption, whereas the PL emission from sample #B may be affected by impurities or other types of defects resulting from intermediate exposure to the ambient environment.

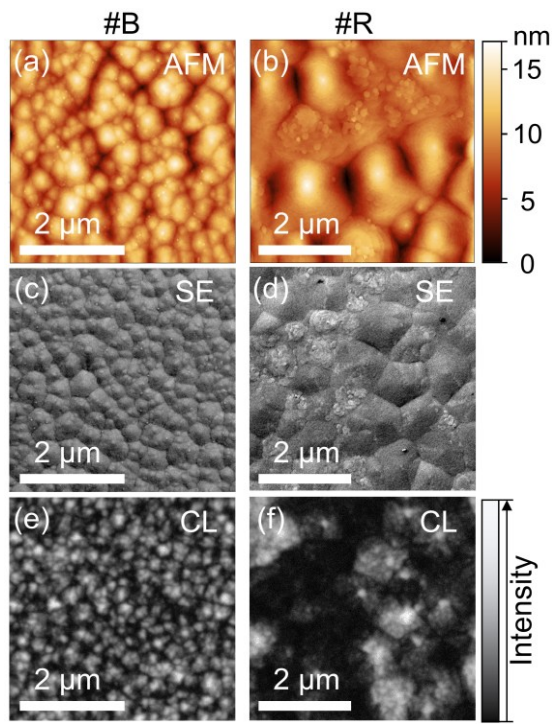

Fig. 3: Comparative analysis of surface morphology and the spatial distribution of light emission. AFM topographs of (a) samples #B and (b) #R with RMS roughnesses of 2.3 nm and 2.1 nm, respectively. Top-view SE micrographs of (c) samples #B and (d) #R. (e) and (f) Panchromatic CL maps corresponding to the areas shown in (c) and (d). The pixels that had <20% of the maximum intensity were considered as dark.

Surface roughness is another critical parameter for pseudo-substrates. Fig. 3(a) and 3(b) show the AFM topographs of samples #B and #R, respectively. Both surfaces are covered with the typical pyramidal mounds dominating the surface morphology of III-N films grown under metal-rich condition.[33,32] However, the mean diameter of the pyramids is about 400 nm on sample #B, compared to approximately 1200 nm for sample #R. This difference is confirmed by the SE micrographs in Fig. 3(c) and 3(d). Despite the different surface morphology, the RSM roughness extracted from the AFM topographs is similar for both samples and amounts to approximately 2 nm.

The spatial distribution of light emission was examined using the panchromatic CL maps in Fig. 3(e) and 3(f). Both maps contain numerous bright features that are roughly circular but irregular in shape. Their average size differs noticeable between the two samples and resembles the size of the pyramids observed on the analogous topo- and micrographs in Fig. 3(a)-(d). Dark regions in the CL maps correspond to non-radiative areas and indicate the presence of defects such as dislocations[34,35] or impurities[36,37]. Typically, on such CL maps, dislocations are associated with well-separated dark spots that can be counted to determine the dislocation density.[35] However, no individial dark spots are observed in the present maps. Assuming a typical spot size corresponding to one dislocation, the integrated dark area in Fig. 3(e) can be estimated to correspond to a high dislocation density in the range $10^9$-$10^{10}$/cm$^2$. At the same time, the dislocation density could also be lower if non-radiative recombination is rather caused by point defects. The precise dislocation density will be determined by transmission electron microscopy analysis of plan-view lamellae. This work is currently ongoing and is not included in the present study. In any case, the comparison of the two CL maps shows that the area fraction of bright regions is higher for sample #B ($\approx$50%) than for sample #R ($\approx$30%). This result directly demonstrates the improvement enabled by our pseudo-substrate growth strategy.

It is not surprising that the CL map of sample #R is mostly dark given that this $In_{0.28}Ga_{0.62}N$ layer was grown directly on a GaN template without any elaborate strategy to accommodate the lattice mismatch of 3%. Relaxation mechanisms in such samples are known to involve the formation of threading dislocations acting as non-radiative recombination centers and mostly crossing the entire layer.[38,39] Indeed, the relaxation degree of this sample is 50%, corresponding to substantial plastic relaxation.

The growth of the pseudo-substrate sample #B involves an intermediate nanostructured surface [cf. Fig. 1(c) and (d)] designed to filter dislocations and/or facilitate elastic relaxation. Indeed, the relaxation degree of this sample is much higher, namely, 80%. At the same time, the non-radiative recombination mechanisms dominant in sample #R may not be completely suppressed. Even assuming that dislocation lines do not propagate from below into the third layer in sample #B, other mechanisms may be relevant. In an effort to elucidate such mechanisms, we consider the size of the relevant features. In principle, the dark areas in Fig. 3(e) could be associated with grain boundary defects formed during the coalescence of misaligned neighboring nanowalls. However, given that the underlying nanowalls grown in step 2 exhibit a mean width of aproximately 30 nm [cf. Fig. 1(e)], each 400-nm-wide nano-pyramid of the coalesced third layer in sample #B must have formed from multiple nanowalls. Speculatively, the presence of extended radiative pyramids could be explained by the scenario that the underlying nanowalls are well aligned or that small misorientations between adjacent nanowalls are elastically accommodated without the formation of grain-boundaries. In this scenario, the non-radiative edges of the islands could correspond to regions where the misorientation between coalescing pyramids is sufficiently large to induce defect nucleation at the boundaries, thereby creating efficient non-radiative recombination centers.[40]

An alternative hypothesis for the dark regions involves the incorporation of impurities, such as Ca, stemming from the deionized water used during substrate cleaning.[36,37] Such impurities can act as deep-level acceptors and show a strong affinity for In atoms, in particular at the low substrate temperatures employed here. Nevertheless, they are sufficiently mobile to migrate along the growth direction and segregate to the surface.[41] Ultimately, they may accumulate at the coalescence boundaries where adjacent nano-pyramids merge and defects form. This process would be analogous to grain-boundary segregation observed in other material systems[42] and would further enhance non-radiative recombination at these boundaries. Similar processes would also be relevant for sample #R grown at the same low substrate temperature. At the same time, in sample #B segregation to the sidewall surfaces of the intermediate (In,Ga)N nanowalls may play an additional role.

In order to investigate more deeply the phenomena relevant for non-radiative recombination and further improve the crystal quality of our pseudo-substrates, we will pursue two different approaches in the follow-up studies: (i) Variation of the starting configuration for coalescence by changing the growth conditions for steps 1 and 2 to influence the intermediate surface morphology of both the rough GaN layer and the (In,Ga)N nanowalls. (ii) Introduction of an (In,Ga)N underlayer to block impurity segregation from the GaN template surface.[41] We note that impurity segregation is likely less problematic for the eventual growth of an LED structure on top of the pseudo-substrate. In particular, such devices are typically grown by metal-organic vapor phase epitaxy (MOVPE) at a higher substrate temperature (700-850 °C), compared to that used for pseudo-substrate growth in PAMBE (500-600 °C). Thus, the impurity incorporation rate is much lower.[36] However, threading dislocations from the underlying layers are likely to propagate into the active region, ultimately degrading device performance. Therefore, quantifying—and reducing, if needed—the dislocation density will be a primary focus of future work.

In comparison to other approaches for the fabrication of (In,Ga)N pseudosubstrates, the RMS surface roughness of ≈2 nm is similar or better than published data for InGaNOS (2.7-3.7 nm),[43] porous nitride underlayers (2-5 nm),[16] and (In,Ga)N layers grown on an (Al,Sc)N buffer (1.6 nm)[44]. Dislocation densities or CL maps are barely available for other types of pseudo-substrates. The combination of an In content of ≈30% with a strain relaxation degree of ≈80% achieved here exceeds the values reported for porous nitride underlayers (In≈0.18, R≈82%),[16] InGaNOS (In≈0.09, R≈55%),[17] and (In,Ga)N pseudo-substrates grown on (Al,Sc)N buffer layers (In≈0.28, R≈20%).[44] Therefore, our three-step (In,Ga)N growth protocol is currently the most promising option to provide the in-plane lattice constant desired for overgrowth of a red (In,Ga)N LED.

In conclusion, we have demonstrated the in-situ fabrication of (In,Ga)N pseudo-substrates exhibiting the target in-plane lattice constant of 3.27 Å and a smooth surface. This PAMBE growth process does not require any complex and costly external processing and may pave the way for the fabrication of highly efficient red (In,Ga)N LEDs.

The authors thank Carsten Stemmler, Hans-Peter Schönherr, and Claudia Herrmann for MBE maintenance and Audrey Gilbert for a critical reading of the manuscript. We also thank Amélie Dussaigne (University of Grenoble-Alpes, CEA, LETI) for the valuable discussions. We appreciate the funding by the German Federal Ministry of Research, Technology, and Space (BMFTR) as well as the Berlin Senate. Additional funding from the European Regional Development Fund thought the application laboratory ZALKAL is gratefully acknowledged.

# Supporting Information

Fabrication of (In,Ga)N pseudo-substrates by a three-step growth protocol without ex-situ processing

Huaide Zhang, Aidan F. Campbell, Jingxuan Kang, Jonas Lähnemann, Oliver Brandt, Lutz Geelhaar [a]

*Paul-Drude-Institut für Festkörperelektronik, Leibniz-Institut im Forschungsverbund Berlin e.V., Hausvogteiplatz 5-7, 10117 Berlin, Germany*

[a] Corresponding author: geelhaar@pdi-berlin.de

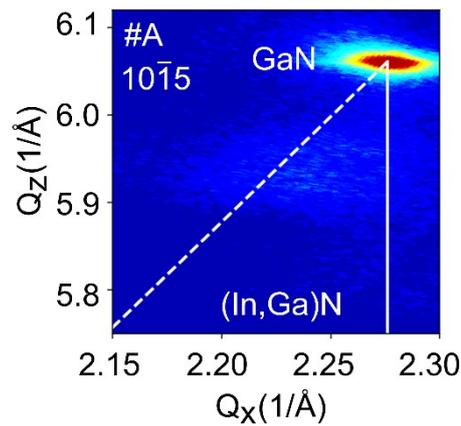

Fig. S1: Reciprocal space map around the $10\bar{1}5$ reflection of sample #A. The position of the (In,Ga)N reflection is diffusive, which may correspond to the gradient In content during the growth and the misalignment of (In,Ga)N nanowalls compared to the underlying GaN template.

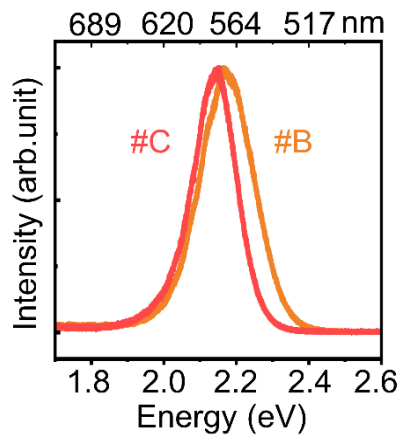

Fig. S2: Room temperature photoluminescence spectra of sample #B and #C. The former spectrum is also presented in Fig. 2(d) of the main document. The latter spectrum exhibits a peak wavelength of 576 nm and a linewidth of 140 meV/37 nm.

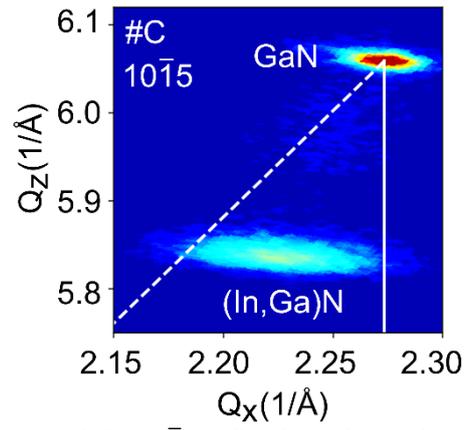

Fig. S3: Reciprocal space map around the 10$\bar{1}$5 reflection of sample #C. The position of the (In,Ga)N reflection corresponds to an In content of ≈32% and a strain relaxation degree of ≈75%.